\documentclass{PoS}%
\usepackage{amsmath}
\usepackage{amsfonts}
\usepackage{amssymb}
\usepackage{graphicx}%
\providecommand{\U}[1]{\protect\rule{.1in}{.1in}}
\title{%
\protect{
\vspace{-1.2cm}
\flushright{
\begin{minipage}{4cm}
\normalsize{
KEK Preprint 2008-36 \\%
CHIBA-EP-173}
\end{minipage}
}
\\%
\vspace{1cm}
}
A new description of lattice Yang-Mils theory and non-Abelian monopoles as the quark confiner%
}

\ShortTitle{A new description of lattice Yang-Mils theory and non-Abelian monopoles as the quark confiner}

\author{\speaker{Akihiro Shibata}\\%
Computing Research Center, High Energy Accelerator Research Organization (KEK) \& \\%
Graduate University for Advanced Studies (Sokendai), Tsukuba 305-0801, Japan\\%
        E-mail: \email{akihiro.shibata@kek.jp}}

\author{Kei-Ichi Kondo\\
Department of Physics, Graduate School of Science, Chiba University, Chiba 263-8522, Japan\\%
        E-mail: \email{kondok@faculty.chiba-u.jp}}
\author{Seikou Kato\\
Takamatsu National College of Technology, Takamatsu 761-8058, Japan\\
        E-mail: \email{kato@takamatsu-nct.ac.jp}}
\author{Shoichi Ito\\
Nagano National College of Technology, 716 Tokuma, Nagano 381-8550, Japan\\
        E-mail: \email{shoichi@ei.nagano-nct.ac.jp}}
\author{Toru Shinohara\\
Graduate School of Science, Chiba University, Chiba 263-8522, Japan\\
        E-mail: \email{inohara@graduate.chiba-u.jp}}

\author{Takeharu Murakami\\
Graduate School of Science, Chiba University, Chiba 263-8522, Japan\\
        E-mail: \email{tom@fullmoon.sakura.ne.jp}}

\abstract{%
 We  propose a new description of the $SU(N)$ Yang-Mills theory on a lattice, which enables one to explain quark confinement based on the dual superconductivity picture in a gauge independent way. This is because we can define gauge-invariant magnetic monopoles which are inherent in the Wilson loop operator.  
For $SU(3)$ there are two options: the minimal option with a single type of non-Abelian magnetic monopole characterized by the maximal stability subgroup $\tilde{H}=U(2)=SU(2)\times U(1)$, 
and the maximal one with two types of Abelian magnetic monopoles characterized by the maximal torus subgroup $\tilde{H}=U(1)\times U(1)$.  The maximal option corresponds to a gauge independent reformulation of the Abelian projection represented by the conventional MAG. 
In the minimal option, we have successfully performed the numerical simulation of the $SU(3)$ Yang-Mills theory on a lattice.
We give preliminary numerical results showing the dominance of the non-Abelian magnetic monopole in the string tension obtained from the Wilson loop in the fundamental representation, and the infrared dominance of a decomposed field variable for   correlation functions after demonstrating the  preservation of color symmetry which was explicitly broken  by the conventional MAG. 

{\bf keywords:} dual superconductivity, quark confinement, monopole dominance 
}

\FullConference{The XXVI International Symposium on Lattice Field Theory\\

		 July 14-19 2008\\
		 Williamsburg, Virginia, USA}

\begin{document}
\section{Introduction}

The dual superconductivity which is believed as the promising mechanism for
quark confinement is conjectured to occur due to the condensation of magnetic
monopoles, just as the ordinary superconductivity is caused by the
condensation of the Cooper pairs\cite{ref:DsuperCond}\cite{ref:tHooft81}. In
the dual superconductor, the dual Meissner effect squeezes the color electric
flux between a quark and an antiquark into a tube like region to form the
hadronic string. The relevant data supporting the validity of this picture
have been accumulated by numerical simulations especially since 1990 and some
of the theoretical predications have been confirmed by these investigations;
the infrared Abelian dominance, magnetic monopole dominance, center vortex
dominance and non-vanishing off-diagonal gluon mass, which are the most
characteristic features for dual superconductivity\cite{ref:EzawaIwasaki}%
\cite{ref:Suzuki}\cite{Greensite}. However, they are confirmed only in
specific gauges such as the maximal Abelian (MA), Maximal Center (MC) and
Laplacian Abelian gauge, which break color symmetry.

We give a new description of the Yang--Mills fields theory on a lattice, which
is expected to give an efficient framework to explain quark confinement based
on the dual superconductivity picture. The description enables us to extract
in a gauge-independent manner the dominant degrees of freedom that are
relevant to quark confinement in the Wilson criterion in such a way that they
reproduce almost all the string tension in the linear inter-quark potential.
We have already given a new framework for the lattice $SU(2)$ Yang--Mills
theory as a lattice version of the Cho-Faddeev-Niemi-Shabanov (CFNS)
decomposition in a continuum theory, and presented numerical evidences for its
validity by performing numerical simulations. \cite{ref:NLCVsu2}%
\cite{ref:NLCVsu2-2}

The issue of generating magnetic monopoles in $SU(N)$ Yang--Mills theory has
been investigated so far under the MAG which breaks the original gauge group
$SU(N)$ into the maximal torus subgroup $H=U(1)^{N-1}$. Then, MAG yields $N-1$
types of Abelian magnetic monopole, in agreement with the observation due to
the homotopy group: $\pi_{2}(SU(N)/U(1)^{N-1})=\pi_{1}(U(1)^{N-1}%
)=\mathbb{Z}^{N-1}$. Therefore, it tends to assume that magnetic monopoles of
$N-1$ types are necessary to cause the dual Meissner effect for realizing
quark confinement. However, it is not yet confirmed whether or not $N-1$ types
of magnetic monopole are necessary to achieve confinement in $SU(N)$
Yang--Mills theory. Rather, we have a conjecture that a single type of
magnetic monopole is sufficient to achieve quark confinement even in $SU(N)$
Yang--Mills theory, once it is defined in a gauge-invariant way. In fact, this
scenario was originally proposed in \cite{kondoShinohara} based on a
non-Abelian Stokes theorem for the Wilson loop operator\cite{KondoNAST}. For
$G=SU(3)$ there are two options: the minimal option with the maximal stability
subgroup $\tilde{H}=U(2)=SU(2)\times U(1)$ is a new one (overlooked so far) on
which we focus in this Talk, while the maximal one with $\tilde{H}=U(1)\times
U(1)$ being equal to the maximal torus subgroup corresponds to a
gauge-independent reformulation of the Abelian projection represented by the
conventional MAG as reported in the lattice 2007 conference\cite{lattce2007}.

\section{New variables on a lattice}

\begin{figure}[ptb]
\begin{center}
\vspace{-3mm} \includegraphics[
width=1.8in, angle=-90
]{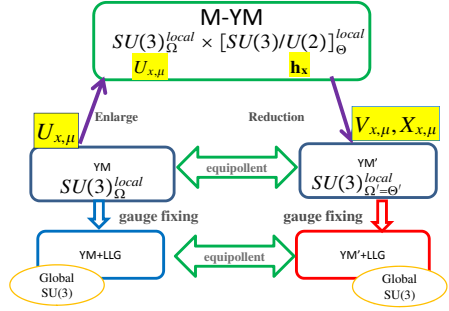} \vspace{-5mm}
\end{center}
\caption{The master Yang-Mills theory: }%
\label{fig:masterYM}%
\end{figure}

First, we summarize the result of a new description of the lattice $SU(N)$
Yang-Mills theory given in Ref.\cite{SCGTKKS08} as an extension of $SU(2)$
case. We wish to construct a lattice formulation in which an ordinary link
variable $U_{x,\mu}$ $\in G=SU(N)$ is decomposed in a gauge-independent manner
into two variables $X_{x,\mu}$ and $V_{x,\mu}$, i.e., $U_{x,\mu}=X_{x,\mu
}V_{x,\mu}$, so that only the variable $V_{x,%
\mu
}$ carries the dominant contribution for quark confinement in agreement with
the dual superconductivity picture. For this purpose, new variables in the
relevant description are supposed to be transformed by a group element
$\Omega_{x}\in G$ as
\begin{subequations}
\begin{align}
U_{x,%
\mu
}  &  \rightarrow U_{x,%
\mu
}^{\prime}=\Omega_{x}U_{x,%
\mu
}\Omega_{x+\mu}^{\dag},\label{eq:GTU}\\
V_{x,%
\mu
}  &  \rightarrow V_{x,%
\mu
}^{\prime}=\Omega_{x}V_{x,%
\mu
}\Omega_{x+\mu}^{\dag},\label{eq:GTV}\\
X_{x,%
\mu
}  &  =X_{x,%
\mu
}^{\prime}=\Omega_{x}X_{x,%
\mu
}\Omega_{x}^{\dag}, \label{eq:GTX}%
\end{align}
where $V_{x,%
\mu
}$ is defined as a link variable and transforms just like the original
Yang-Mills link variable $U_{x,\mu},$ while $X_{x,%
\mu
}$ is defined like a site variable representing a matter field and transforms
according to the adjoint representation. It is important to see that
decomposed variables are required to be transformed by the full $SU(3)$ gauge
group. In the conventional Abelian projection, on the contrary, the $V_{x,\mu
}$ field is identified with an Abelian part and is supposed to be transformed
as an Abelian field.

The relationship between lattice variables and gauge variables in continuum
theory \cite{kondoShinohara} are given by%
\end{subequations}
\begin{subequations}
\begin{align}
U_{x,%
\mu
}  &  =X_{x,%
\mu
}V_{x,%
\mu
}=\exp\left(  -ig\int dx^{\mu}\mathbf{A}_{\mu}(x)\right)  ,\label{var:U}\\
V_{x,%
\mu
}  &  =\exp\left(  -ig\int dx^{\mu}\mathbf{V}_{\mu}(x)\right)  =\exp\left(
-ig\epsilon\mathbf{V}_{\mu}(x+\epsilon\mu/2)\right)  ,\label{var:V}\\
X_{x,%
\mu
}  &  =\exp\left(  -ig\epsilon\mathbf{X}_{\mu}(x\right)  ). \label{var:X}%
\end{align}

In order to obtain the new variables by the decomposition respecting the gauge
transformation property given in the above, we consider the extended
Yang-Mills theory, called the master Yang-Mills theory (see Figure
\ref{fig:masterYM}), by introducing a \textit{single} type of the color
field:
\end{subequations}
\begin{equation}
\mathbf{h}_{x}:=\Theta_{x}\mathrm{diag}(1/\sqrt{3},1/\sqrt{3},-2/\sqrt
{3})\Theta_{x}^{\dag}\in SU(3)/U(2)=G/\tilde{H},
\end{equation}
such that it transforms according to the adjoint representation under an
independent gauge transformation $\Theta_{x}\in G=SU(3).$ The decomposition
$U=VX$ is determined by solving the defining equation;
\begin{subequations}
\begin{align}
&  D_{\mu}^{\epsilon}[V]\mathbf{h}_{x}:=\frac{1}{\epsilon}\left(  V_{x,\mu
}\mathbf{h}_{x+\mu}-\mathbf{h}_{x}V_{x,\mu}\right)  =0,\label{Defeq1}\\
&  \mathrm{tr}(X_{x,\mu}\mathbf{h}_{x})=0, \label{DefEq2}%
\end{align}
where eq.(\ref{Defeq1}) represents that $\mathbf{h}(x)$ is covariantly
constant in the background $V_{x,\mu}$ and eq.(\ref{DefEq2}) means that
$X_{x,\mu}$ has the vanishing $\tilde{H}$-commutative part. \footnote{It is
shown \cite{KondoShibata} using a non-Abelian Stokes theorem that a set of
defining equations is obtained as a necessary and sufficient condition for the
Wilson loop operator to be dominated by the decomposed variable $V_{x,\mu}$ in
the sense $W_{C}[U_{x,\mu}]\cong($const.$)W_{C}[V_{x,\mu}].$} The solution of
the defining equations are obtained uniquely for given $U_{x,\mu}$ and
$\mathbf{h}_{x}$ by way of a newly defined variable $\tilde{V}_{x,\mu}$ which
does not belong to $SU(3)$:
\end{subequations}
\begin{equation}
\tilde{V}_{x,\mu}:=U_{x,\mu}+\frac{2\sqrt{3}}{5}\left(  \mathbf{h}_{x}%
U_{x,\mu}+U_{x,\mu}\mathbf{h}_{x+\mu}\right)  +\frac{24}{5}\mathbf{h}%
_{x}U_{x,\mu}\mathbf{h}_{x+\mu}.
\end{equation}
In fact, the $SU(3)$ variable $V_{x,\mu}$ is obtained using the polar
decomposition, together with another $SU(3)$ variable $X_{x,\mu}$ as%
\begin{equation}
V_{x,\mu}=\underline{V_{x,\mu}}\left(  \det\underline{V_{x,\mu}}\right)
^{-1/3},\quad\underline{V_{x,\mu}}:=\left(  \tilde{V}_{x,\mu}\tilde{V}_{x,\mu
}^{\dag}\right)  ^{-1/2}\tilde{V}_{x,\mu}\text{ ,}\qquad X_{x,u}=U_{x,\mu
}V_{x,\mu}^{\dag}.
\end{equation}
In order to obtain the equipollent theory (written in terms of the new
variables) with the original Yang-Mills theory, we impose the reduction
condition which plays the role of reducing the extended gauge symmetry
$SU(3)_{\Omega}\times\left[  SU(3)/U(2)\right]  _{\Theta}$ \ to the original
gauge symmetry $SU(3)_{\Omega=\Theta}$ so that the new variables $V_{x,\mu
},X_{x,\mu}$ and $\mathbf{h}_{x}$ transform under the same group
$\Omega=\Theta\in SU(3)$.
Such a reduction condition can be given by minimizing the functional $F_{RC}$
of $U_{x,\mu}$ and $\mathbf{h}_{x}$ under the independent local gauge
transformations \{$\Omega,\Theta$\}:
\begin{equation}
F_{RC}[\Omega,\Theta;U_{x,\mu},\mathbf{h}_{x}]=\sum_{x,\mu}\mathrm{tr}\left(
(D_{\mu}^{\epsilon}[^{\Omega}U_{x,\mu}]^{\Theta}\mathbf{h}_{x})(D_{\mu
}^{\epsilon}[^{\Omega}U_{x,\mu}]^{\Theta}\mathbf{h}_{x})^{\dag}\right)  .
\label{eq:;RC}%
\end{equation}
By definition, the reduction condition (\ref{eq:;RC}) is invariant under the
gauge transformation $\Theta=\Omega$ and does not break the original gauge
symmetry as expected. Therefore, we can impose any gauge fixing afterwards.

\section{Numerical simulations}

Next, we proceed to the numerical simulations. The link variable
configurations $\{ U_{x,\mu}\}$ can be generated by using the standard method,
since the decomposition $U_{x,\mu}=V_{x,\mu}X_{x,\mu}$ is done in a
gauge-invariant manner. The color field $\mathbf{h}_{x}$ is determined so as
to fulfill the reduction condition using the same algorithm as that used in
the the gauge fixing procedure. We have generated the link variable
configurations for the Wilson lattice action using $16^{4}$ lattice at
$\beta=5.70$, and obtained the new variables by imposing the lattice Landau
gauge (LLG) for the original Yang-Mills field, rather than the MA gauge which
breaks color symmetry explicitly.

\begin{figure}[ptb]
\begin{center}
\includegraphics[
height=1.9in,
]{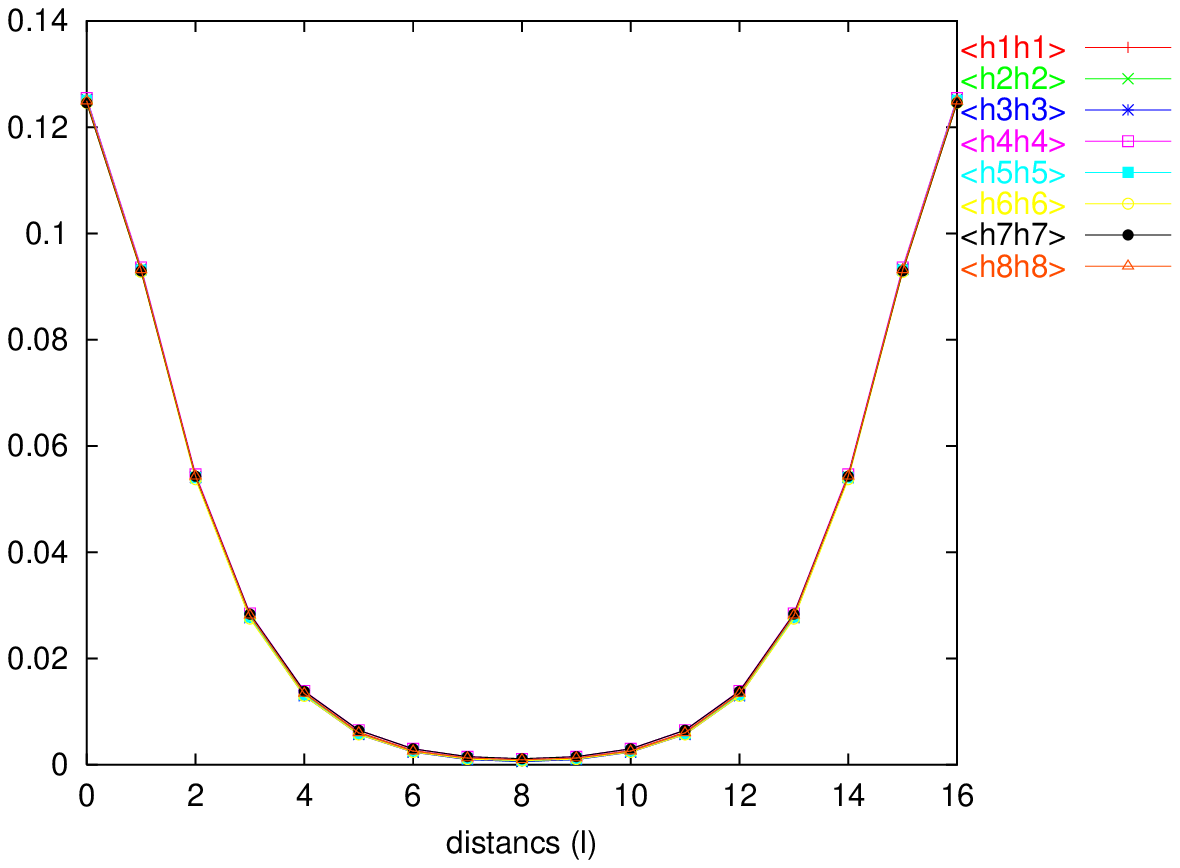}\includegraphics[
height=1.9in
]{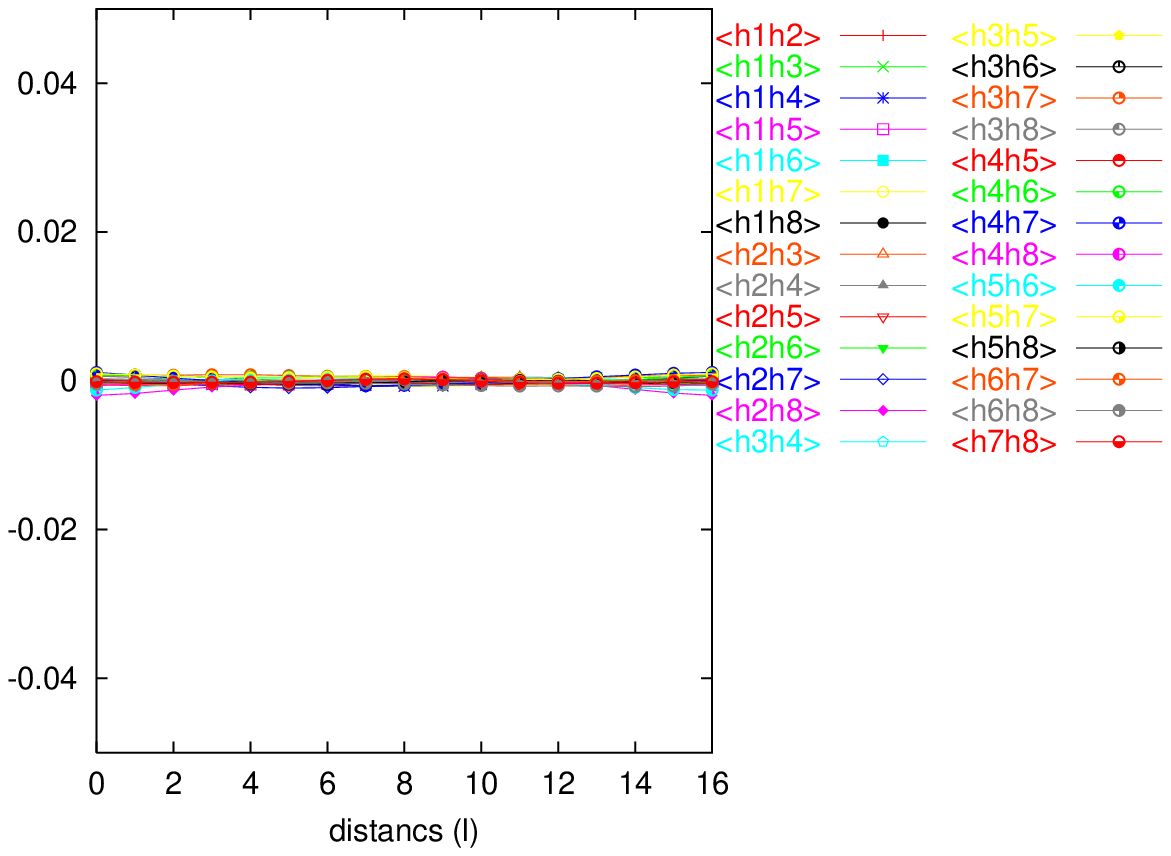}
\end{center}
\caption{The correlation functions of the color field. (Left panel) diagonal
parts, (Right panel) off-diagonal parts.}%
\label{fig:color-symm}%
\end{figure}

We first focus on the color field. The new variables are defined so as to
preserve the local gauge symmetry of the original Yang-Mills theory. Even
after imposing LLG, therefore, the Yang-Mills theory should respect the global
gauge symmetry, i.e., the color symmetry. This issue can be tested by
measuring the vacuum expectation value (VEV) of the color field $\mathbf{h}%
_{x}$ and its correlation functions. The VEV of a color field shows
$\left\langle h^{A}\right\rangle =0$ for $\ A=1,2,..,N^{2}-1=8$. In Figure
\ref{fig:color-symm}, we give a plot of correlation functions $\left\langle
h^{A}(x)h^{B}(0)\right\rangle $ for $A,B=1,...,8$. All the diagonal parts
$\left\langle h^{A}(x)h^{A}(0)\right\rangle $ have the same non-vanishing
correlations, while all the off-diagonal parts $\left\langle h^{A}%
(x)h^{B}(0)\right\rangle $ ($A\neq B$) are vanishing. Thus, the correlation
functions is of the form
\begin{equation}
\left\langle h^{A}(x)h^{B}(0)\right\rangle =\delta^{A,B}D(x),
\end{equation}
which implies that the color symmetry is preserved. 

\begin{figure}[ptb]
\begin{center}
\includegraphics[
height=1.7in,
]{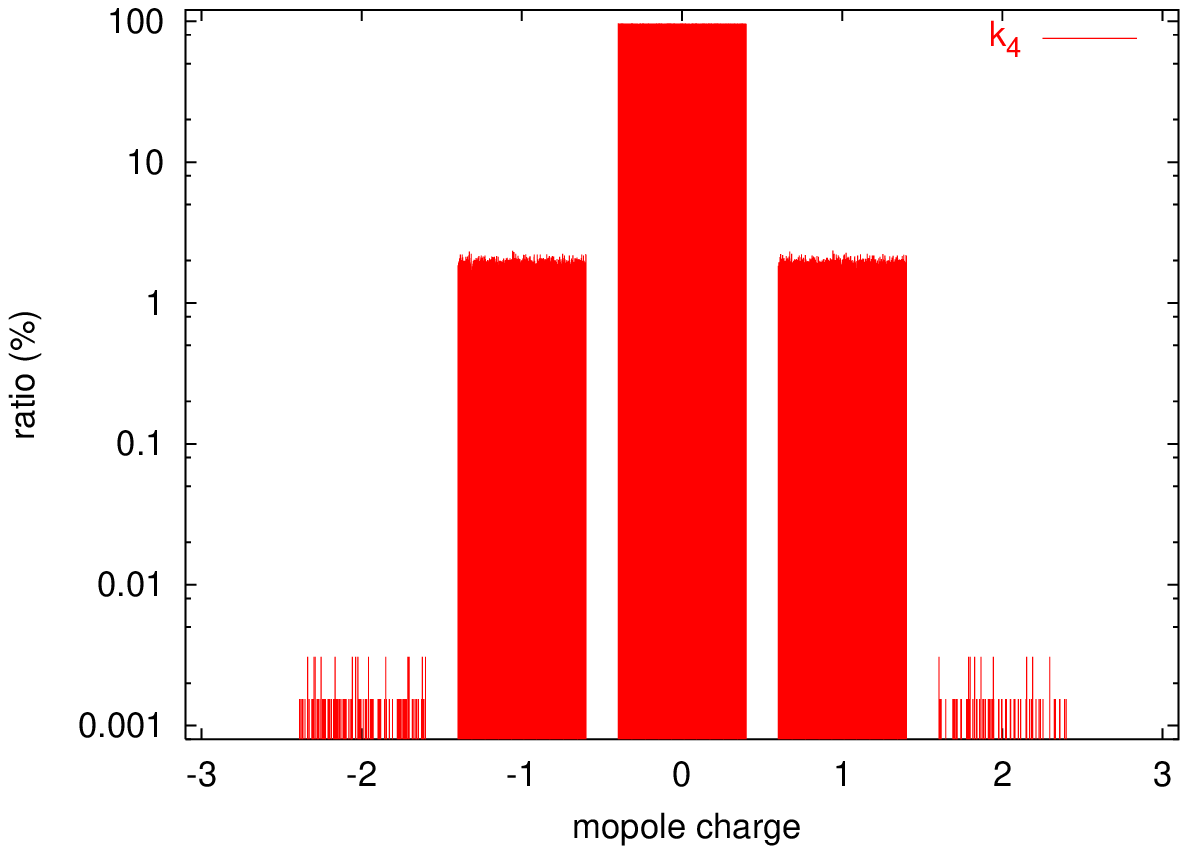}\includegraphics[
height=1.7in,
]{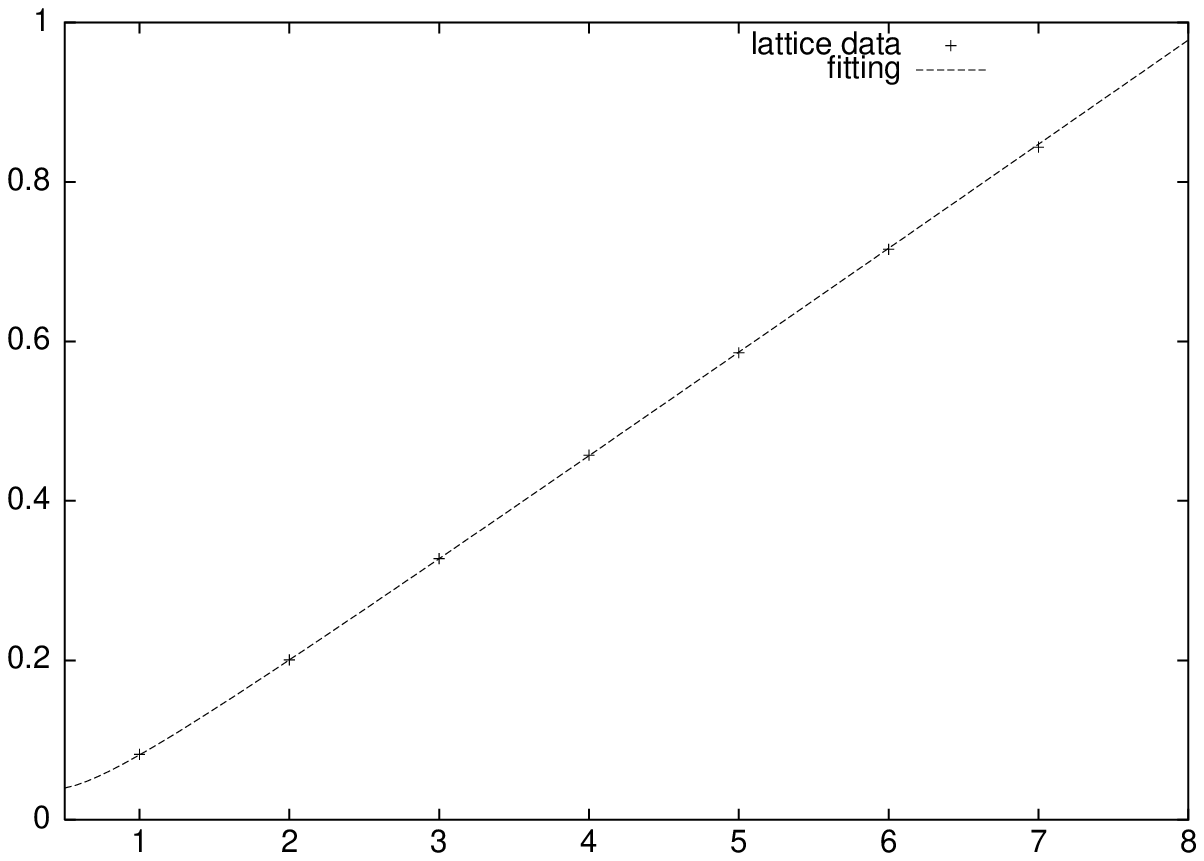}
\end{center}
\caption{(Left panel) the magnetic monopole charge distribution, (Right panel)
the static interquark potential calculated only from the magnetic monopole
part.}%
\label{fig:monopole}%
\end{figure}

Next we investigate the static interquark potential through the new variables.
The Wilson loop operator in the fundamental representation in the continuum
theory is rewritten into \cite{KondoNAST}
\begin{align}
W_{c}[\mathbf{A}]  &  =\mathrm{tr}\left[  P\exp ig%
{\displaystyle\oint_{C}}
\mathbf{A}_{\mu}(x)dx^{\mu}\right]  /\mathrm{tr}(\mathbf{1})=\int d\mu
\lbrack\xi]_{\Sigma}\exp\left\{  ig\int_{\Sigma:\partial\Sigma=C}dS^{\mu\nu
}F_{\mu\nu}[\mathbf{V}]\right\} \nonumber\\
&  =\int d\mu\lbrack\xi]_{\Sigma}\exp\left\{  ig(K,\Xi_{\Sigma}%
)+ig(J,N_{\Sigma})\right\}  , \label{eq:NAST-c}%
\end{align}
where we have defined $K:=\delta{}^{\ast}F$, $J:=\delta F$, $\Xi_{\Sigma
}:=\delta{}^{\ast}\Theta_{\Sigma}\Delta^{-1}$ and $N_{\Sigma}:=\delta
\Theta_{\Sigma}\Delta^{-1}$ with the four-dimensional Laplacian,
$\Delta=d\delta+\delta d$. Here $\Theta_{\Sigma}$ is the vorticity tensor
defined by $\Theta_{\Sigma}^{\mu\nu}=\int_{\Sigma}dS^{\mu\nu}(X(\sigma
))\delta(x-X(\sigma))$ on the surface $\Sigma:\partial\Sigma=C$ spanned by the
Wilson loop C. Note that $\mathbf{V}_{\mu}(x)$ field in eq(\ref{eq:NAST-c}) is
equivalent to the $\mathbf{V}_{\mu}(x)$ reproduced from the new variable
$V_{x,\mu}$ in the continuum limit. Therefore, the magnetic monopole
contribution to the Wilson loop average on a lattice is given by%
\begin{subequations}
\begin{align}
&  \left\langle W_{C}[U]\right\rangle \cong\left\langle W_{C}[V]\right\rangle
=\left\langle \exp\left(  i\sum_{x,\mu}K_{x,\mu}\Xi_{x,\mu}\right)
\right\rangle ,\label{eq;WLmon}\\
&  \Xi_{x,\mu}=%
{\displaystyle\sum_{s}}
\Delta^{-1}(s-s^{\prime})\frac{1}{2}\epsilon_{\mu\alpha\beta\gamma}%
\partial_{\alpha}S_{\beta\gamma}^{J}(s+\mu),\text{ }\partial_{\alpha}%
S_{\alpha\beta}^{J}=J_{\beta},
\end{align}
with the lattice magnetic monopole current $K_{x,\mu}$ given by%
\end{subequations}
\begin{equation}
K_{x,\mu}:=\frac{1}{2}\epsilon_{\mu\lambda\alpha\beta}\partial_{\lambda}%
\Theta_{x,\alpha\beta}^{8}[V],\quad\Theta_{a\beta}^{8}[V]:=-\arg
\mathrm{tr}\left(  (\frac{1}{3}\mathbf{1-}\frac{2}{\sqrt{3}}h_{x})V_{x,\alpha
}V_{x+\alpha,\beta}V_{x+\beta,\alpha}^{\dag}V_{x,\beta}^{\dag}\right)  ,
\end{equation}
where $\partial_{\lambda}$ denotes the forward difference (lattice derivative)
in the $\lambda$ direction: $\partial_{\lambda}f(x):=f(x+\epsilon\hat{\lambda
})-f(x)$. It should be noticed that the magnetic monopole current $K_{x,\mu}$
is gauge invariant, as can be seen from the transformation law of the new
variables.
This monopole should be identified with a \textquotedblleft
non-Abelian\textquotedblright\ magnetic monopole, since the monopole current
is defined from $V_{x,\mu}$ which involves $\mathbf{h}_{x}\in G/\tilde
{H}=SU(3)/U(2)$. The left panel of Figure \ref{fig:monopole} shows the
distribution of the magnetic charge of the relevant magnetic monopoles
obtained from 200 configurations, which have integral quantized values:
$n_{x,\mu}=K_{x,\mu}/2\pi\in\mathbb{Z}$. The right panel of Figure
\ref{fig:monopole} show the static potential calculated from the monopole part
according to eq(\ref{eq;WLmon}). The numerical data of the static potential
$V_{m}(R)$ extracted from $\left\langle \exp\left(  i\sum_{x,\mu}K_{x,\mu}%
\Xi_{x,\mu}\right)  \right\rangle $ is well fitted by a function
$V_{m}(R)=-\alpha_{m}/R+\sigma_{m}R$ with the value $\sigma_{m}=0.13301(36)$.
In comparison with the full string tension, for example, the result in
Ref.\cite{ref:R.G.Edward}, $\sigma_{full}=0.3879(39)$, thus, we have shown the
non-Abelian magnetic monopole dominance for the string tension in the $SU(3)$
Yang-Mills theory:
\begin{equation}
\sigma_{m}/\sigma_{full}=0.87\pm0.19.
\end{equation}

\bigskip

\begin{figure}[ptb]
\begin{center}
\includegraphics[
height=1.8in,
]{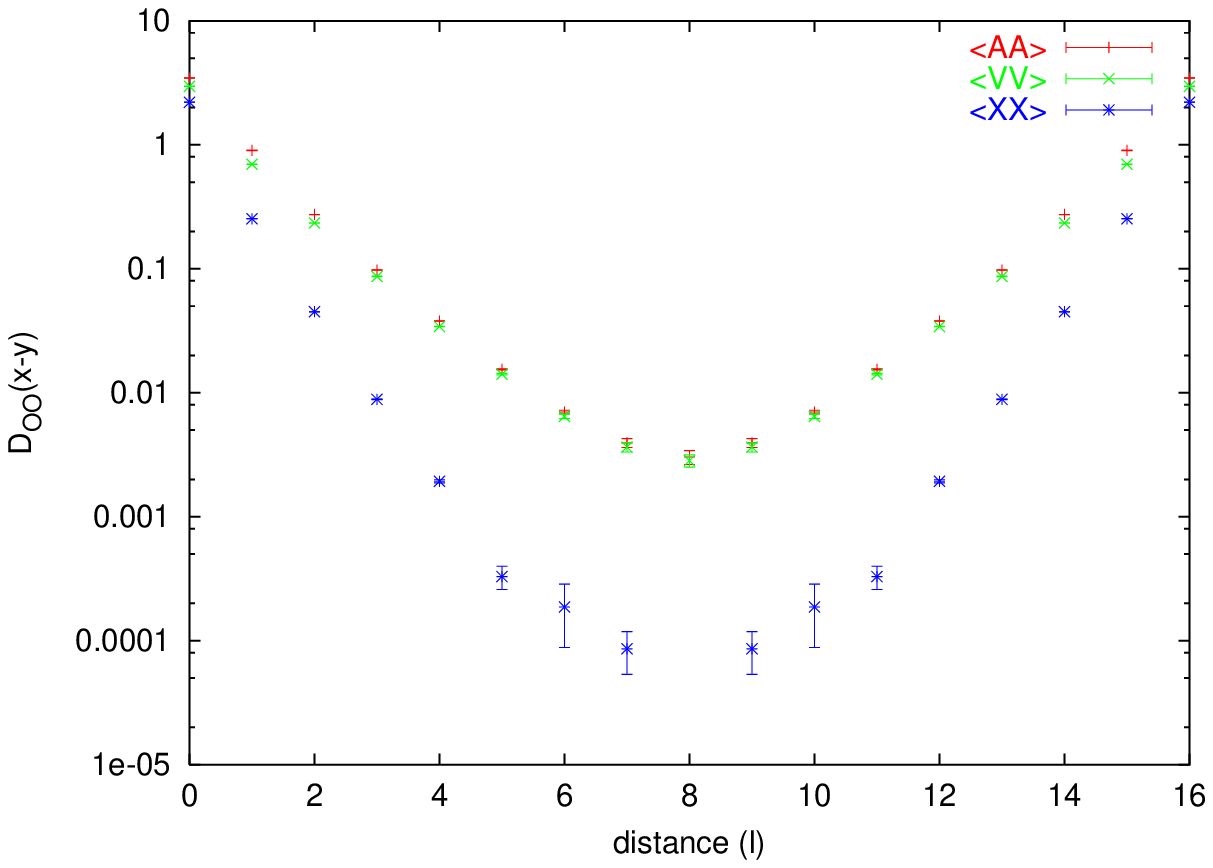}\includegraphics[
height=1.8in
]{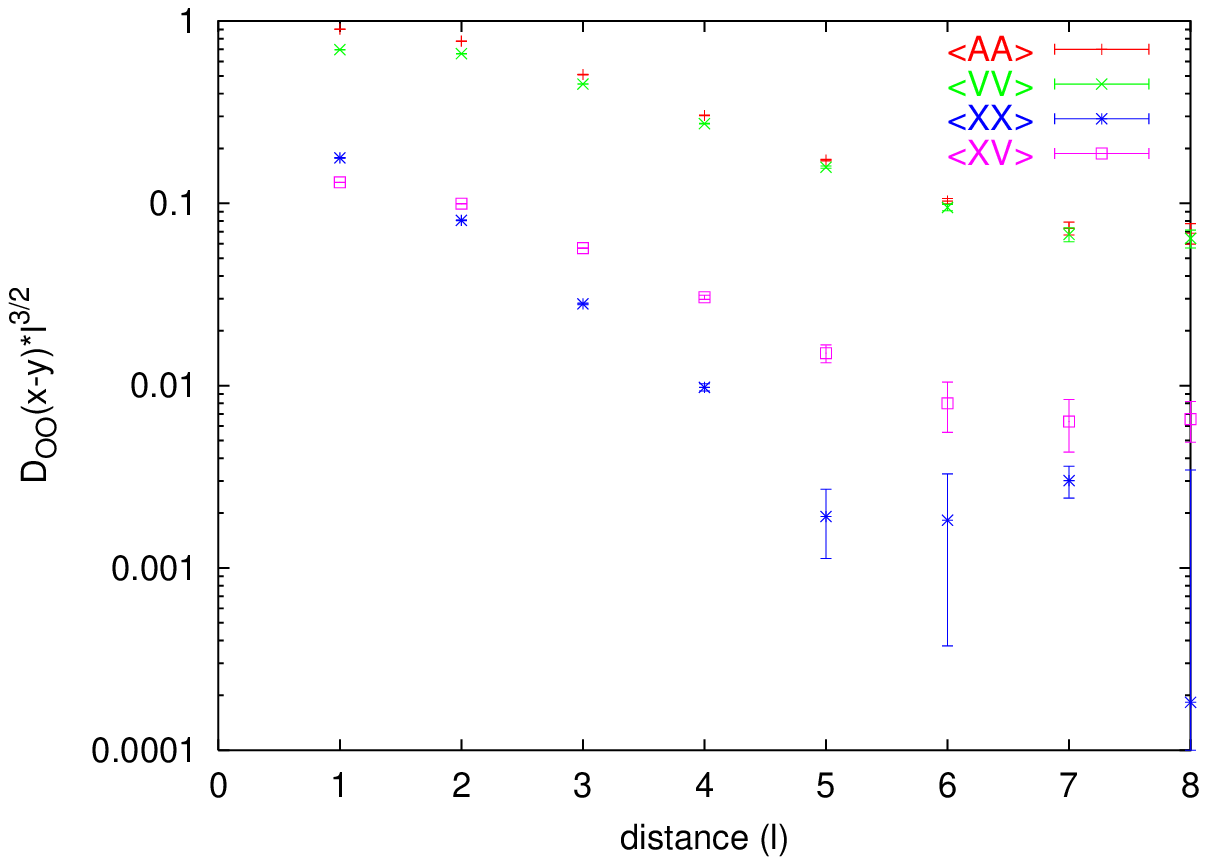}
\end{center}
\caption{The correlation functions of original gauge field and new variables.
(Left panel) The logplot of correlation functions. (Right panel) The scaled
plot.}%
\label{fig:propagator}%
\end{figure}

Finally,\ we devote to the correlation functions. The 2-point correlation
functions (propagators) of the new variables and the original Yang-Mills gauge
potential are defined by
\begin{equation}
D_{OO}(x-y):=\left\langle O_{\mu}^{A}(x)O_{\mu}^{A}(y)\right\rangle \text{ for
}O_{\mu}^{A}(y^{\prime})\in\{\mathbf{V}_{x^{\prime},\mu},\mathbf{X}%
_{x^{\prime},\mu},\mathbf{A}_{x^{\prime},\mu}\},
\end{equation}
where an operator $O_{\mu}^{A}(x)$ is defined by the linear type, e.g.,
$\mathbf{A}_{x^{\prime},\mu}:=(U_{x,\mu}-U_{x,\mu}^{\dag})/2g\epsilon$. The
left panel of Figure \ref{fig:propagator} shows the behavior of correlation
functions, $D_{AA}$, $D_{VV}$ and $D_{XX}$ vs. the distance $l=|x-y|$. The
correlation function $D_{VV}$ damps slowly and has almost the same damping as
$D_{AA}$, while the $D_{XX}$ damps quickly. This result suggests that the $V$
part of the gluon propagator is dominated in the infrared region and mass
generation by $X$ part. As the variable $X_{x,\mu}$ transforms as an adjoint
matter, (see eq.(\ref{eq:GTX})), the Yang-Mills theory can have a
gauge-invariant mass term $\mathcal{L}_{M_{X}}=M_{X}^{2}\mathrm{tr}%
(\mathbf{X}_{\mu}^{2})$. The Fourier transformation of the massive propagator
behaves for large $M_{X}r$ as
\begin{equation}
D_{XX}(r)=\int\frac{d^{4}k}{(2\pi)^{4}}e^{ik(x-y)}\frac{3}{k^{2}+M_{x}^{2}%
}\simeq\frac{3\sqrt{M_{X}}}{2(2\pi)^{3/2}}\frac{e^{-M_{X}r}}{r^{3/2}},
\end{equation}
and hence the scaled propagator $r^{3/2}D_{XX}(r)$ should be proportional to
$\exp(-M_{X}r).$ The right panel of Figure \ref{fig:propagator} shows the
logarithmic plot of the scaled propagators in LLG. This suggests the mass
generation of gluons, although there exists the mixing between $V$ and $X$ in
LLG and $M_{X}$ cannot be identified with the mass of $X$ straightforwardly.

\section{Summary and discussions}

We have given the new description of the lattice Yang-Mills theory to give a
gauge independent decomposition of the link variable, $U_{x,\mu}=X_{x,\mu
}V_{x,\mu}$, which approves the the gauge independent formulation of the dual
superconductivity picture. We have performed the numerical simulation in the
minimal case of the $SU(3)$ lattice Yang-Mills theory, and have demonstrated
the color symmetry restoration, the non-Abelian monopole dominance in a gauge
invariant way and the infrared $V$ dominance in LLG other than MAG. We have
also shown based on a non-Abelian Stokes theorem that the string tension for
the fundamental quark is explained by the non-Abelian magnetic monopole
defined from $V$ part in the minimal case.

To establish the dual superconductivity picture by the new description of the
lattice Yang-Mills theory, we need further study for the full $V$ dominance,
e.g., ``electric'' and ``magnetic'' from the $V$ part reproduces the full
string tension, the N-ality property, and so on. These subjects are under
investigation and will be discussed in a separate paper.

%

\section*{Acknowledgement}%

This work is financially supported by Grant-in-Aid for Scientific Research (C)
18540251 from Japan Society for the Promotion of Science (JSPS) and the Large
Scale Simulation Program No. 08-16 (FY2008) of High Energy Accelerator
Research Organization (KEK).

%


\end{document}